\documentstyle[12pt,psfig]{article}
\begin{document}
\baselineskip .3in
\begin{titlepage}
\begin{center}
{\Large {\bf Effect of Fibonacci modulation on superconductivity}}
\vskip .2in
{\em  Sanjay Gupta$^{\dag}$, Shreekantha Sil$^{\ddag}$ and 
Bibhas Bhattacharyya$^{\dag \dag}$}

\vskip .1in
$^{\dag}$Department of Theoretical Physics, Indian Association for the Cultivation
of Science, Kolkata -- 700032, India.\\
$^{\ddag}$Department of Physics, Vishwabharati, Shantiniketan 731235, Birbhum, West Bengal, INDIA\\
$^{\dag \dag}$Department of Physics, Scottish Church College,
1 \& 3, Urquhart Square, Kolkata -- 700006, INDIA

\vskip .3in
{\bf Abstract}
\end{center}
\noindent
We have studied finite-sized single band models with short range pairing 
interactions between electrons in
presence of diagonal Fibonacci modulation in one dimension. Two models, namely 
the attractive
Hubbard model and the Penson-Kolb model, have been investigated at half-filling
at zero temperature by solving the Bogoliubov-de Gennes equations in real space
within a mean field approximation. The  competition between ``disorder'' and
the pairing interaction leads to a suppression of superconductivity (of
usual pairs with zero centre-of-mass momenta) in the 
strong-coupling limit while an enhancement of the pairing correlation is 
observed in the weak-coupling regime for both the models. However, 
the dissimilarity
of the pairing mechanisms in these two models brings about notable difference 
in the results.
The extent to which the bond ordered wave and the $\eta$-paired (of pairs with 
centre-of-mass momenta = $\pi$) phases of the Penson-Kolb model
are affected by the disorder has also been studied in the present 
calculation. 
Some finite size effects are also identified.  
\end{titlepage}
{\bf I. Introduction}

The competition between electronic correlation and disorder remains one of the
prime issues of investigation in condensed matter physics during the last few
years \cite{Gulac, Gia, Hida, GSS1}.
However, the effect of such disorder is yet to be fully
explored in the context of superconductivity. Some earlier
 experiments
showed that in case of some weak coupling
superconductors, $T_c$  increases with increasing disorder
strength, while for some strong coupling materials 
 $T_c$ is nearly insensitive to the  strength of disorder
\cite{buck, berg, comb}.
On the other hand very recent experiments observed \cite{goldman}
destruction of superconductivity 
with increasing disorder strength in some low dimensional
superconductors. 
Experiments studying the effects of disorder on 
superconducting A-15 materials have shown that in these substances
$T_c$ decreases with
increasing disorder \cite{Strong, Sweed,  ghosh}.
 Theoretical investigation of properties of 
superconductors in presence of disorder was addressed by
Anderson \cite{ander} way back in 1959. He showed that the presence
of weak non-magnetic impurities, does not 
suppress superconductivity appreciably.
 Quite recently
 \cite{ran} the effect of bulk impurity on superconductivity
has been  
studied within the negative-$U$ Hubbard model with random disorder in the
site potentials.
It showed that superconductivity is suppressed by disorder. Moreover, 
it also showed that disorder introduces spatial inhomogeneity in
pairing correlation.
Models like the negative-$U$ Hubbard model, which give rise to 
 superconductivity within a short range pairing mechanism,
have been studied extensively in the context of high $T_c$ and
other exotic superconductors \cite{robask}.
Therefore, the studies of interplay between disorder and
 superconductivity within the framework of short range 
pairing mechanism would be of great interest. In this  work we describe a 
study on the effect of Fibonacci-modulated disorder on two different
 models of superconductivity
which support short range pairing. 

The models of short ranged pairing that we have focused on in the present work are the negative-$U$ Hubbard model \cite{robask} and the Penson-Kolb (PK) model 
\cite{penson} respectively. These two models have been extensively studied 
\cite{robask, penson, capone, nathan, marsiglio, loh, Achille, bhatt, hartmann, japaridze, bulka} in the recent past owing to their tentative relevance in the field of high-T$_c$ cuprates and organic superconductors. However, the pairing mechanism of the two models are of different physical origin. The negative-$U$ Hubbard model supports a short ranged pairing due to an on-site attraction in sharp 
contrast with the non-local pairing mechanism generated by a pair hopping process in the PK model. Such a non-local pair-hopping mechanism gives
rise to a very rich phase diagram of the PK model \cite{Achille,bhatt,
hartmann,japaridze,bulka} as compared to the case of the negative-$U$ Hubbard model. Thus a comparative study of these two models in presence of ``disorder'' is expected to reveal a qualitative difference in the nature of competition between disorder and pairing correlation.

The present study has been restricted to one dimension (1-d) because of the 
possibility of checking the present results against the earlier studies on the 
two models, the PK model in particular, by a variety of techniques in 1-d.
At least in the asymptotic cases, some of rigorous results are always 
available in 1-d rather in higher dimensions. This led us to choose a typical 
1-d model of diagonal aperiodicity, namely, a Fibonacci modulated sequence of 
the site potentials, for observing the effect of disorder on these superconducting 
models. This type of quasi-crystalline ``disorder'' not only interpolates between 
the extreme cases of full grown order and random disorder, but also qualifies for 
the scope of experimental investigations owing to the availability of various 
quasi-crystalline superlattices in recent times \cite{merlin}. Thus our main 
objective, in 
this paper, is to understand the qualitative manner in which the quasi-crystalline 
disorder modifies the superconducting correlation in two specific models with 
different pairing mechanism and how does the nature of this competition change 
from the weak- to the strong-coupling limit.

 Our investigations for both the models concentrate on  
 decoupling of the Hamiltonian within a mean field approximation
(MFA) followed by a self consistent solution of
 the Bogoliubov-de Gennes (BdG)
 equations in real space \cite{ran} for the decoupled Hamiltonians.
The use of a mean-field approach is usually questionable in low dimensions. However, even in low dimensional system, such a technique works satisfactorily in a broken-symmetry phase \cite{GSB}. The two models of short-ranged pairing that we have studied here are known to exhibit several broken-symmetry phases 
in the ordered limit. Previous mean-field calculations \cite{ robask, bulka} 
captured these phases satisfactorily in low dimensions and were found to compare
 well with the results obtained by other methods \cite{loh, bhatt, hartmann}. Moreover, in the present scheme of calculation relevant parameters are determined self-consistently for each and every site separately which captures successfully the spatial fluctuations induced by the Fibonacci disorder \cite{GSS1}. In fact the present scheme of calculation has
already been tested with success for a randomly disordered Hubbard model in 2-d \cite{ran}.

In section II  we  define  the Hamiltonians and their generic features.
 In section III
we furnish the BdG equations. Section IV deals with
 results that have been obtained in the present calculation.
 Section V summarizes this work. 
\\ \\
\noindent
{\bf II.  The models}
\\
\noindent
{\bf (a) The negative-$U$ Hubbard model:}

The negative-$U$ (attractive) Hubbard Hamiltonian with Fibonacci modulation in the site 
potentials is given by:
\begin{equation}
{\cal H}_{-U}=\sum_i (\epsilon_i-\mu) n_i - (t\sum_{i\sigma}
 c^{\dag}_{i\sigma} c_{i+1\sigma}
+ h.c.) + U \sum_i c^{\dag}_{i\uparrow} c_{i\uparrow} c^{\dag}_{i\downarrow} c_{i\downarrow},
\end{equation}
where, $c^{\dag}_{i\sigma} (c_{i\sigma})$ creates (destroys) an electron of spin 
$\sigma (\sigma=\uparrow,\downarrow)$ on the $i$-th site. $n_{i\sigma}=
c^{\dag}_{i\sigma} c_{i\sigma}$ and $n_i = n_{i\uparrow}+n_{i\downarrow}$.
$\epsilon_i$ is the site energy at the $i$-th site;
 it takes on the value $\epsilon_A$ or $\epsilon_B$
according to the Fibonacci sequence: $ABAABABAABAABABAABAAB .....$;  $\mu$ is the
chemical potential and $t$ is the single particle hopping integral. The last term
is the on-site Hubbard interaction. We will take negative
values of $U$ in our calculations for the attractive Hubbard model.

The attractive Hubbard model without any modulation in site potential
($\epsilon_i=0$) has been extensively studied
 \cite{robask, capone, nathan, marsiglio,
pedersen}. In this limit ($\epsilon_i=0$), there is a competition
between the single-particle hopping ($t$) and the Hubbard correlation ($U < 0$). The on-site Hubbard correlation favours formation of localized singlet pairs of electrons while 
the hopping term tends to break the pairs. Due to the local pairing mechanism
 superconducting (SC) state
 and charge density wave (CDW) phases become
degenerate in the ground state for a half-filled
 band at zero temperature \cite{loh}.   
This model has been extensively used in describing high-$T_c$ and other
related superconducting systems \cite{robask}. Effect of random diagonal disorder on the attractive Hubbard interaction has already been studied \cite{ran, Mic}. 
It is interesting to observe how the
Fibonacci modulation alters its superconducting properties.
\noindent
\\
\noindent
{\bf (b) The Penson-Kolb model}

The PK model with Fibonacci-modulated site potentials 
is written below:
\begin{equation}
{\cal H}_{PK}=\sum_i (\epsilon_i-\mu) n_i - (t\sum_{i\sigma} c^{\dag}_{i\sigma} c_{{i+1} 
\sigma} + h.c.) - V( \sum_i c^{\dag}_{i+1\uparrow} c^{\dag}_{i+1\downarrow} c_{i\downarrow} c_{i\uparrow} + h.c.)
\end{equation}

The first two terms have similar implications as in the Hubbard Hamiltonian (1).
The third term is the pair hopping term which is responsible for transfer of a
 singlet pair of electrons ($\uparrow\downarrow$) between neighbouring
sites. $V$ is the nearest neighbour pair hopping amplitude.
 PK model with $\epsilon_i=0$,
 favours formation of singlet pairs in real space due to this 
short range pair hopping and therefore, shows a non-local pairing mechanism.
In this sense the study of this model is complementary to the study 
of the on-site pairing model, namely the negative-$U$ Hubbard model.
The PK model ($\epsilon_i=0$) 
and its various generalizations have been widely studied over the years
   \cite{Achille,bhatt,hartmann,japaridze,bulka}.
 The ground state phase diagram of this model
for a half-filled band in one dimension is understood to a great extent
as it stands at present. For $V > 0$, this model shows superconducting
instability which corresponds to usual pairing with pairs of zero
 centre of mass momenta.
However for $V < 0$, in the strong coupling limit (beyond some value of $V$,
say $V_c$), this model is shown to
exhibit $\eta$-pairing with center of mass momentum $q=\pi$
 \cite{bhatt,
japaridze, bulka}. 
There is a phase which does not support SC ordering for $V_c < V < 0$. 
A real space RG calculation showed the existence of a CDW phase
in this regime \cite{bhatt} and did not
consider the possibility of the existence of antiferromagnetic
bond order wave (BOW) which was later
shown to coexist with the CDW within a Bosonization calculation
\cite{hartmann}.
A mean field study, however, showed only the antiferro-BOW state 
in this region \cite{bulka}.
It seems rather interesting how these properties are modified in
 presence of disorder. We focus our attention to a Fibonacci modulation in 
$\epsilon_i$ as in the case of the negative-$U$ Hubbard model.
\\ \\
\noindent
{\bf III.  The calculation within the MFA and the BdG equations}
\\ 
\noindent
{\bf (a) For the negative-$U$ Hubbard model with Fibonacci modulation}

 We solve
the negative-$U$ Hubbard model first by decoupling the Hamiltonian
in favour of superconductivity and then by solving 
the BdG equations of motion in real space in a self consistent manner.
The decoupled Hamiltonian looks like:
\noindent
\begin{eqnarray}
{\cal H}_{-U}&=&\sum_i (\epsilon_i-\mu) n_i -(t\sum_{i\sigma} c^{\dag}_{i\sigma}
c_{i+1\sigma} + h.c.) + U \sum_i <n_{i\uparrow}> c^{\dag}_{i\downarrow} c_{i\downarrow}\nonumber\\&&+ U \sum_i <n_{i\downarrow}> c^{\dag}_{i\uparrow} c_{i\uparrow}-U \sum_i <n_{i\uparrow}><n_{i\downarrow}>+
 U\sum_i <c^{\dag}_{i\uparrow} c^{\dag}_{i\downarrow}> c_{i\downarrow} c_{i\uparrow}\nonumber\\
&&+U\sum_i <c_{i\downarrow} c_{i\uparrow}> c^{\dag}_{i\uparrow} c^{\dag}_{i\downarrow}
-U \sum_i <c^{\dag}_{i\uparrow} c^{\dag}_{i\downarrow}><c_{i\downarrow} c_{i\uparrow}>
\end{eqnarray}

The BdG equations of motion for the operators
$c_{i\uparrow}$ and $c^{\dag}_{i\downarrow}$ are: 
\begin{equation}
i\dot{c}^{\dag}_{i\uparrow}=[c^{\dag}_{i \uparrow}, {\cal H}_{-U} ] = tc^{\dag}_{i-1\uparrow}
 + tc^{\dag}_{i+1\uparrow}-U\Delta^{\dag}_i c_{i\downarrow}
-U<n_{i\downarrow}> c^{\dag}_{i\uparrow}
-(\epsilon_{i}-\mu) c^{\dag}_{i\uparrow}
\end{equation}
\begin{equation}
i\dot{c}_{i\downarrow}=[c_{i \downarrow}, {\cal H}_{-U} ]= -tc^{\dag}_{i+1\downarrow} - tc^{\dag}_{i-1\downarrow}-U\Delta_{\i} c^{\dag}_{i\uparrow}+U<n_{i\uparrow}> c_{i\downarrow}+(\epsilon_{i}-\mu) c_{i\downarrow}
\end{equation}
where 
$\dot{c}_{i\sigma}=\frac{dc_{i\sigma}}{dt}$
and
$\Delta^{\dag}_{i}= <c^{\dag}_{i\uparrow} c^{\dag}_{i\downarrow}>$. We study the Fourier 
transform of the superconducting gap parameter as defined by,
\begin{equation}
\Delta_{q}=\left(1/N \right)\sum_{i\sigma} e^{iq.r_{i}} \Delta_{i}
\end{equation}
where, $r_i$ is the position of the $i$-th site.
\\
\noindent
{\bf (b) For the Penson-Kolb model with Fibonacci modulation}

The decoupled PK Hamiltonian is given below:
\begin{eqnarray}
{\cal H}_{PK}&=&\sum_i (\epsilon_i-\mu) n_i - \left[(t+V p^{\dag}_{i\downarrow})\sum_{i} c^{\dag}_{i\uparrow} c_{i+1
\uparrow} + h.c.\right]-\left[(t+V p^{\dag}_{i\uparrow})\sum_{i} c^{\dag}_{i\downarrow} c_{i+1
\downarrow} + h.c.\right]\nonumber\\
&&-V( \sum_i \Delta_i c^{\dag}_{i+1\uparrow} c^{\dag}_{i+1\downarrow}+ h.c.)-V( \sum_i \Delta_{i+1} c^{\dag}_{i\uparrow} c^{\dag}_{i\downarrow}+h.c.)
+(V\Delta^{\dag}_i\Delta_{i+1}+c.c.)\nonumber\\
&&+(V p^{\dag}_{i\downarrow} p^{\dag}_{i\uparrow}+c.c.)
\end{eqnarray}
where 
$\Delta^{\dag}_{i}= <c^{\dag}_{i\uparrow} c^{\dag}_{i\downarrow}>$
and
$p^{\dag}_{i\sigma}= <c^{\dag}_{i+1\sigma} c_{i\sigma}>$.

 The BdG equations corresponding to the PK Hamiltonian
for the operators $c_{i\uparrow}$ and $c^{\dag}_{i\downarrow}$ in
real space are given below:
\noindent
\begin{eqnarray}
&&i\dot{c}^{\dag}_{i\uparrow}=(t+V p^{\dag}_{i\downarrow}) c^{\dag}_{i-1\uparrow}+(t+V p_{i\downarrow}) c^{\dag}_{i+1\uparrow}+V\Delta^{\dag}_{\i-1} c_{i\downarrow}+V\Delta^{\dag}_{i+1} c_{i\downarrow}-(\epsilon_{i}-\mu) c^{\dag}_{i\uparrow}
\\
&&i\dot{c}_{i\downarrow}= -(t+V p^{\dag}_{i\uparrow}) c^{\dag}_{i+1\downarrow} - (t+Vp_{i-1\uparrow}) c^{\dag}_{i-1\downarrow}+V\Delta_{\i+1} c^{\dag}_{i\uparrow}+V\Delta_{i-1} c^{\dag}_{i\uparrow}+(\epsilon_{i}-\mu) c_{i\downarrow}
\end{eqnarray}

We solve the BdG equations in a self consistent fashion to determine the
Fourier transform of the superconducting gap (6) together with
the  bond order parameter in $q$-space as given by:
\begin{equation}
B_{q}=\left(1/2N \right) \sum_{i \sigma} \sigma e^{iq.r_{i}} p_{i\sigma}~.
\end{equation}
\\ \\
\noindent
{\bf IV.  Results}
\\
\noindent
{\bf (a) For the Fibonacci-modulated negative-$U$ Hubbard model}

 The negative-$U$ Hubbard model has been studied at half-filling on a
 one dimensional
chain for system sizes $N$=34 and 144. We have chosen $\epsilon_A=0$ always.
The plot of $\Delta_q$ in Fig.1 
reveals a maximum at $q=0$.
This bears a signature of normal superconducting phase with singlet pairs having zero centre of mass momentum.  
The nature of competition between the disorder and correlation strongly
depends on the value of $U$. This can be seen from the plots of $\Delta_0$
 against $U$ (negative values) in
 Fig.s 2(a) and 2(b) for $N=34$ 
 and 144 respectively.
It is observed that for lower values of $|U|$, value of $\Delta_0$
is enhanced due to Fibonacci modulation. 
This is due to formation of double occupancies at the sites of
lower energy. 
 A crossover takes place at intermediate values of $|U|$. 
After that $\Delta_0$ decreases with increasing Fibonacci modulation.
This happens due to increased backscattering
in presence of Fibonacci ``disorder''.
 Another important fact to note is that the value of $|U|$
 where the crossover takes place becomes lower with increasing
 system size; this is a reflection of stronger effect of electronic correlation
 in larger system sizes. 
Fig.3 shows the plot of $\Delta_i$ against $i$ at $U=-2.8$
 (a value of $|U|$ below which the crossover has taken place) for $N$=34.
 This plot clearly
depicts the spatial inhomogeneity in pairing brought about by
 the Fibonacci modulation.
Such results are in agreement with \cite{ran}.
In the periodic limit the distribution of $\Delta_i$
in space is more or less 
homogeneous except for the boundary effects.
In presence of disorder $\Delta_i$ closely follows the Fibonacci pattern.
Competition between disorder and correlation is further revealed
in Fig.4 showing the plot of $\Delta_0$ as a function of
 $\epsilon_B$. 
It is interesting to observe that for low values of $|U|$ ($U=-1.0$),
 $\Delta_0$ increases with Fibonacci
disorder while for strong $|U|$ ($U=-4.0$)  $\Delta_0$ sensibly
decreases with 
increasing disorder. This is precisely what we have observed in Fig.2.
\\
\noindent
{\bf (b) For the Fibonacci-modulated  Penson-Kolb model}

The one dimensional PK model also has been studied for half-filling
and for the same system sizes ($N$=34 and 144). Let us first discuss 
the case of $V > 0$. 
 Fig.s 5(a) and 5(b) show the plots of 
$\Delta_{q}$ for $V$=1.0. 
The peaks of $\Delta_{q}$ at $q=0$ for $V > 0$ reveal that the
system is in a normal superconducting phase. 
 The effect of Fibonacci disorder
for the high and low values of $V$ can be seen from  Fig. 6, the plot of 
$\Delta_0$  against $\epsilon_B$ at $V=1.0$
 and $V=2.0$ respectively for $N=34$. 
For low values of $V$, 
 disorder makes the value of $\Delta_0$ increase
while the opposite phenomenon occurs  at larger values of $V$. 
However, the fall in $\Delta_0$ with $\epsilon_B$ for
 $V=2.0$ is extremely slow compared to the notable variation of $\Delta_0$ 
for $V=1.0$. 
Thus, even a very strong disorder
does not seriously affect the superconducting properties of the system 
at high values of $V$. This is in sharp contrast to the case of large $|U|$.
This can be qualitatively understood in the following way. A closer look into the 
eqn.s (4) and (5) reveals that the Hubbard interaction directly renormalizes the 
``effective'' site potentials for the electrons of opposite spins to different 
extents. Thus for the larger values of $|U|$ the site potential seen by an up-spin electron
at a particular site differs considerably from that seen by a down-spin electron. Therefore,
the possibility of formation of singlet pairs is reduced due to the competition between
strong correlation and disorder. But this should not be so drastic in case
of the PK model as 
suggested by eqn.s (8) and (9).
Let us now discuss the variation of the superconducting gap $\Delta_0$
as a function of the pair hopping
amplitude $V$ (for its positive values).
Fig.s 7(a) and 7(b) are the corresponding plots for $N$=34 and 144 
respectively at three different values of the site energy $\epsilon_B$=0, 0.4
and 1.0 at half-filling.
 In both the plots it is observed that $\Delta_0$ remains vanishingly small
until
a value of $V$, say $V_0$, is reached. The value of $V_0$ 
decreases with system size. The reason behind this is that the finite size gap
in the energy spectrum is larger than the gap in the spectrum
 due to $V$, for $V < V_0$.
So the system is unable to realize the effect of $V$ in this region and hence 
$\Delta_0$  remains almost zero.
 For $N=144$ the finite size gap reduces; consequently the
value of $V_0$ goes down. For $V > V_0$ , $\Delta_0$ increases smoothly
and sharply with $V$.
 The interesting point to note is that both in Fig.s 7(a) and 7(b), 
$\Delta_0$ has higher values for
 $\epsilon_B \neq 0$ (compared to the case of $\epsilon =0$) in the
regime of small $V$. 
In this region the effect of pair hopping
 is rather weak to form a large number of singlet pairs. However, 
the Fibonacci modulation generates sites of lower energy which favour
formation
of ``doublons''.
 As a result, a higher value of $\Delta$
is observed for $\epsilon_B\neq 0$ than for  $\epsilon_B=0$.
 A crossover takes place at a certain $V$ after
which the entire 
 process reverses.
For large values of V, pair hopping
process can generate a large number of pairs. But the pairing
 becomes suppressed in the presence of the Fibonacci modulation. 
This is qualitatively similar to the negative-$U$ case. 
However, the degree of suppression
of $\Delta_0$ due to disorder in the PK model is not very high. 
Consistent with this observation we also find that the spatial
 inhomogeneity in $\Delta_i$ is less 
pronounced in case of the PK model (Fig.8) as compared to the negative-$U$ model. 

Next we consider the case $V < 0$. We study the superconducting properties
 for larger values of $|V|$.
The plots of $\Delta_{q}$ against $q$ show  peaks 
at $q=\pi$ in Fig.s 9(a) and 9(b)
 suggesting that the system is in $\eta$-paired (with centre of mass
momentum $q=\pi$) superconducting
phase $\cite{bhatt}$. The peaks are suppressed by Fibonacci aperiodicity.
In Fig.s 10(a) and 10(b) we show the plots of $\Delta_{\pi}$ against $V$
in the regime $V < -2$ for $N$=34 and $N$=144.
It is found that for $V =-2.0$, $\Delta_{\pi}$ has higher
value in the periodic limit ($\epsilon_B=0$) for both $N=34$ and $N=144$
than in the cases of $\epsilon_B$=0.4 and 1.0.
It is interesting to note that the suppression of $\eta$-pairing due
to disorder becomes less pronounced in larger system size. Consequently
the decrease of the $\eta$-SC correlation across $|V|\approx 2.0$ is
rather gradual (Fig. 10(b))with increasing disorder in larger systems.

Next we discuss the intermediate region of $-2 < V < 0$. In this regime 
our result shows a bond-order wave (BOW) phase \cite{hartmann, bulka}
in the periodic limit (for $|V|<1.9$)  
which is evident from the plot of $B_q$ against $q$ (Fig.11).
 The bond order
parameter shows peaks at $q=\pi$ for $V=-1.0$ in Fig.s 11(a) and 11(b) for
$N=34$ and $N=144$ respectively.
Interestingly enough, the introduction
 of aperiodicity enhances the peak
 in  case of $N$=34, whereas
for $N$=144, the peak in the periodic limit is a little bit higher
than that in the Fibonacci-modulated case.
 This appears to be a finite size effect. 
The height of the peak (i.e. the BOW) is found to increase  with increasing 
$|V|$ (below a certain $|V_c| \approx 1.9$ above which
$\eta$-pairing takes place). 
Fig.s 12 and 13 depict simultaneous variations of the BOW order parameter and 
the $\eta$-paired superconducting gap parameter as functions of the
 pair hopping interaction
$V$ in the regime $-2 < V < 0$. 
It is observed in Fig.s 12(a) and 13(a) that in the periodic limit 
the transition from the BOW phase to the $\eta$-paired phase occurs at around
$|V_c|\approx 1.9$ for both $N$=34 and 144. This value of $|V_c|$ 
increases with increasing Fibonacci modulation in case of $N$=34.
But this does not happen for the case $N$=144 as observed in Fig.13.
This is a result of the fact that Fibonacci sequence can 
significantly disrupt and delay the
formation of $\eta$-paired ordering for smaller system sizes, a fact already
apprehended in Fig.s 10(a) and 10(b). 
For smaller system sizes the finite size gaps are large enough to mask the
effects of the pair hopping which reduces the BOW in the region
$-2 < V < 0$. Therefore the quasiperiodic disorder makes the value of
$|V_c|$ shift in an appreciable manner. For $N$=144, the finite size gap
becomes much smaller and the effect of the pair hopping dominates. Therefore, 
$|V_c|$ becomes insensitive to the effect of the disorder.
\\ \\
\noindent
{\bf IV. Conclusion}

Summarizing, we have studied the finite-sized negative-$U$ Hubbard and
Penson-Kolb models
with Fibonacci-modulated site potentials at 
 half filling in one dimension. In the present work,
 we used a mean field superconducting decoupling followed by 
self consistent solutions of
the corresponding BdG equations in real space. 
We have calculated
the superconducting gap parameters  
for normal and $\eta$-paired phases and the
BOW order parameter for the antiferromagnetic bond-ordered phase
in case of the PK model. 
For the negative-$U$ Hubbard model we obtain the normal
superconducting phase corresponding to centre of mass momentum $q=0$.
Here we observe a crossover at intermediate values of
$|U|$. Below this crossover the pairing correlation increases due to Fibonacci 
disorder while above it the pairing is suppressed by disorder.
In case of the PK model we observe a similar crossover
in the positive $V$ sector.
However the suppression of SC correlation due to ``disorder'' is much reduced
in the PK model as compared to the negative-$U$ Hubbard model.
In case of the negative-$U$ Hubbard model the real space pairing amplitude
shows a spatial inhomogeneity which closely follows the underlying
quasiperiodic pattern while the feature is less prominent in case of 
the PK model. The apparent reason behind this (as we have discussed
earlier) can be understood
from the difference in the nature of competition between aperiodicity
and pairing mechanism in these two models. The non-local pairing mechanism
in the PK model is already known to give results which are 
qualitatively different from the negative-$U$ Hubbard model \cite{ranroybh}.
However, in the present study, the qualitative 
difference between these two has been further clarified via the mechanism 
of competition between disorder and pairing term.
In the $\eta$-paired phase of the PK model,
 superconductivity is suppressed by disorder in general.  
In identifying the bond ordered phase our calculations match with
predictions of previous calculations \cite{hartmann, bulka}. 
It turns out that the transition point from BOW to $\eta$-SC phase
is not severely affected by the presence of disorder in large systems.

The present study, thus, reveals the qualitative feature of the competition 
between Fibonacci disorder and short ranged pairing interactions within the 
framework of the negative-$U$ Hubbard model and the PK model. It is interesting 
to note how the microscopic mechanism  of pairing modifies the nature of 
competition in going from the weak- to the strong-coupling regime. It is to
be noted here that these set of results do not show marked variation in going 
from $N=34$ to $N=144$. On the other hand, the finite size effects that have 
been identified in the present calculation are essentially controlled by (as 
discussed in Sec. IV) the finite size gap which goes roughly as $1/N$. Thus no dramatic change in these results are expected to take place beyond $N=144$.
 It may, however, be noted that for obtaining very precise quantitative estimate 
of quantities, e.g. the transition point from the BOW to the $\eta$-SC phase in the PK model, a finite size 
scaling would be necessary. Some future work may proceed in this direction. 

For further investigation in this scheme,
 it would be interesting to include the repulsive Hubbard interaction
term in the PK model and see the effect of  
quasiperiodic modulations in such systems.
 Also a detailed
 study of the effect of 
band filling away from the half filled case would be very much interesting.
The possibility of formation of a CDW \cite{bhatt, hartmann}, which
is left out in the present calculation, could also be investigated.
It would also be significant to study the case of finite temperatures
and higher dimensions. The effect of random diagonal disorder in the
PK model also needs some attention.

\newpage
\noindent
{\bf Figure Captions}
\\
\\
\noindent
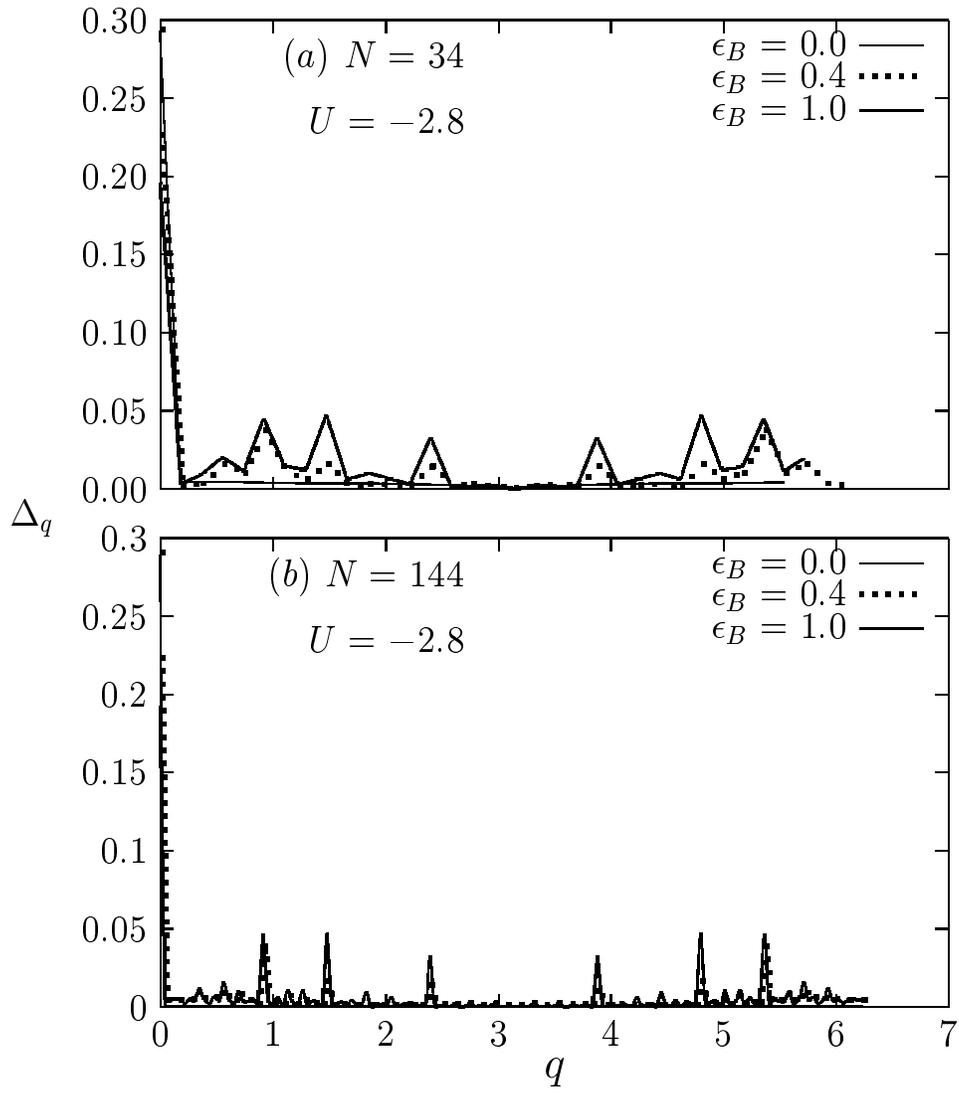
Fig.1. Plots of $\Delta_q$, the Fourier transform of the superconducting gap
parameter, as a function of $q$ for different ``disorder'' strengths ($\epsilon_B$) in the 
Fibonacci-modulated attractive Hubbard model (for $U = -2.8$) for
(a) $N=34$ and (b) $N=144$. Scale of energy is given by $t=1.0$.
\\
\\
\noindent
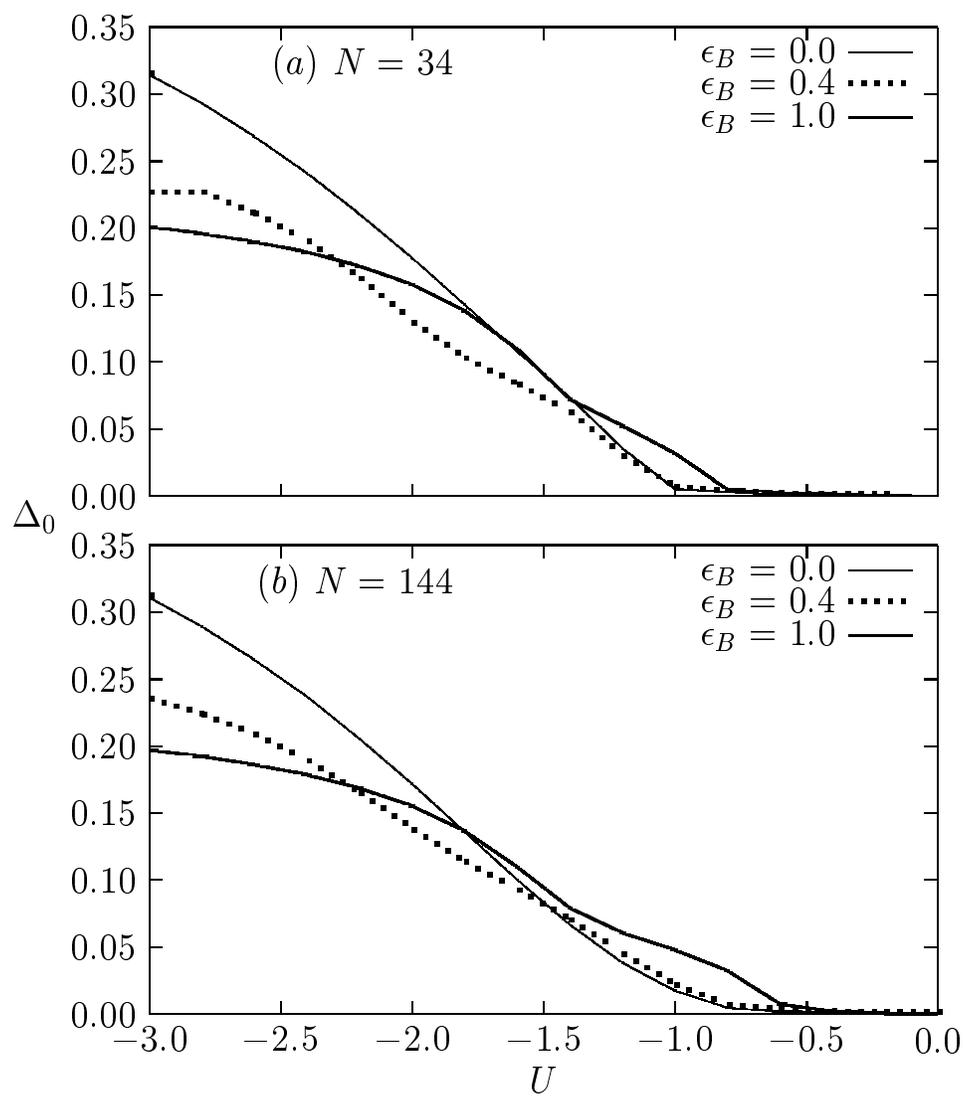
Fig.2. Plots of the superconducting gap parameter $\Delta_0$ (for pairs with zero
centre-of-mass momentum) vs. $U$ in the Fibonacci-modulated attractive Hubbard 
model for (a) $N=34$ and (b) $N=144$ for different ``disorder'' strengths
($\epsilon_B$). Scale of energy is given by $t=1.0$.
\\
\\
\noindent
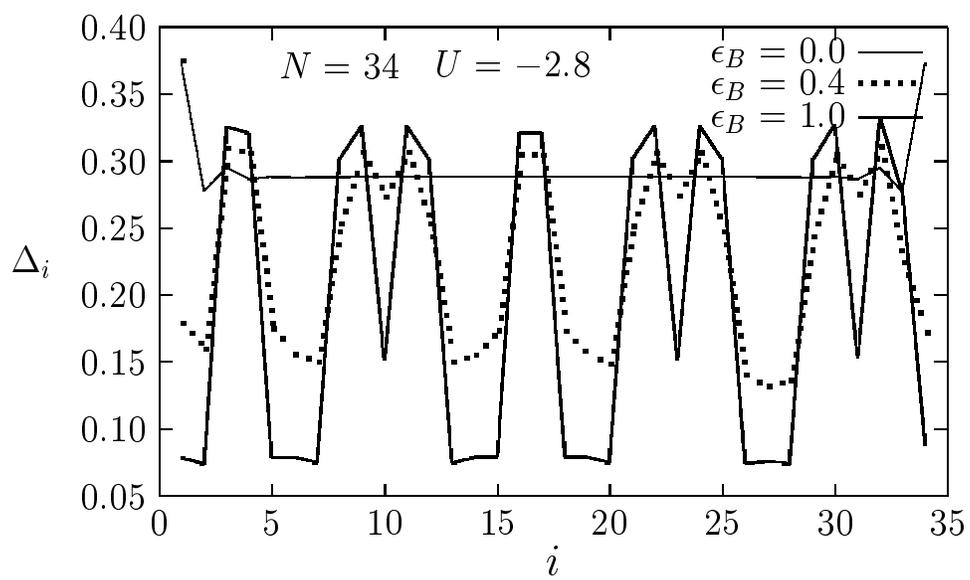
Fig.3. Plots of the local pairing amplitude $\Delta_i$ vs. site index $i$ for
the attractive Hubbard model with different strengths of Fibonacci modulation.
The plot clearly reveals an underlying Fibonacci pattern which is
strengthened for larger values of $\epsilon_B$. Scale of energy is given by $t=1.0$.
\\
\\
\noindent
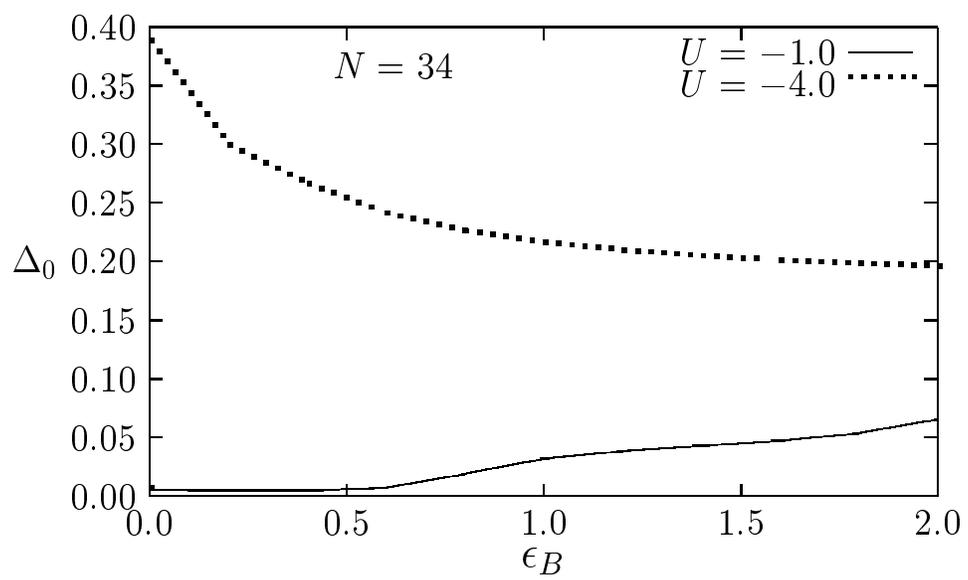
Fig.4. Plots of $\Delta_0$ vs. $\epsilon_B$ in the Fibonacci-modulated 
attractive Hubbard model for different values of $U$. The seemingly different 
behaviour for smaller and larger values of $U$ can be understood from the 
crossover noted in Fig.2. Scale of energy is given by $t=1.0$.
\\
\\
\noindent
Fig.5. Plots of $\Delta_q$ vs. $q$ in the Fibonacci-modulated Penson-Kolb
model for the positive $V$ vector for (a) $N=34$ and (b) $N=144$ for different
strengths of ``disorder'' ($\epsilon_B$) (scale of energy is given by $t=1.0$).
\\
\\
\noindent
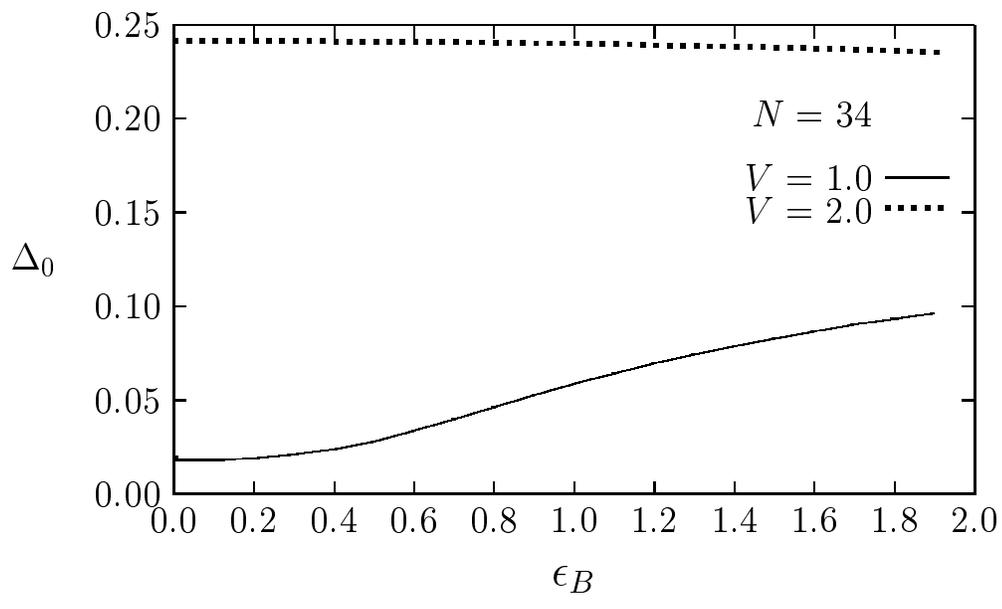
Fig.6.  Plots of $\Delta_0$ vs. $\epsilon_B$ in the Fibonacci-modulated 
Penson-Kolb model for different values of pair hopping interaction ($V$) in
the positive $V$ sector. Effect of ``disorder'' is found to be distinctly
different for low and high values of $V$. However, the fall of $\Delta_0$
in the strong-coupling limit (e.g. for $v=2.0$) is notably slower compared to
the case of the attractive Hubbard model (Fig.4) (scale of energy is given by $t=1.0$).
\\
\\
\noindent
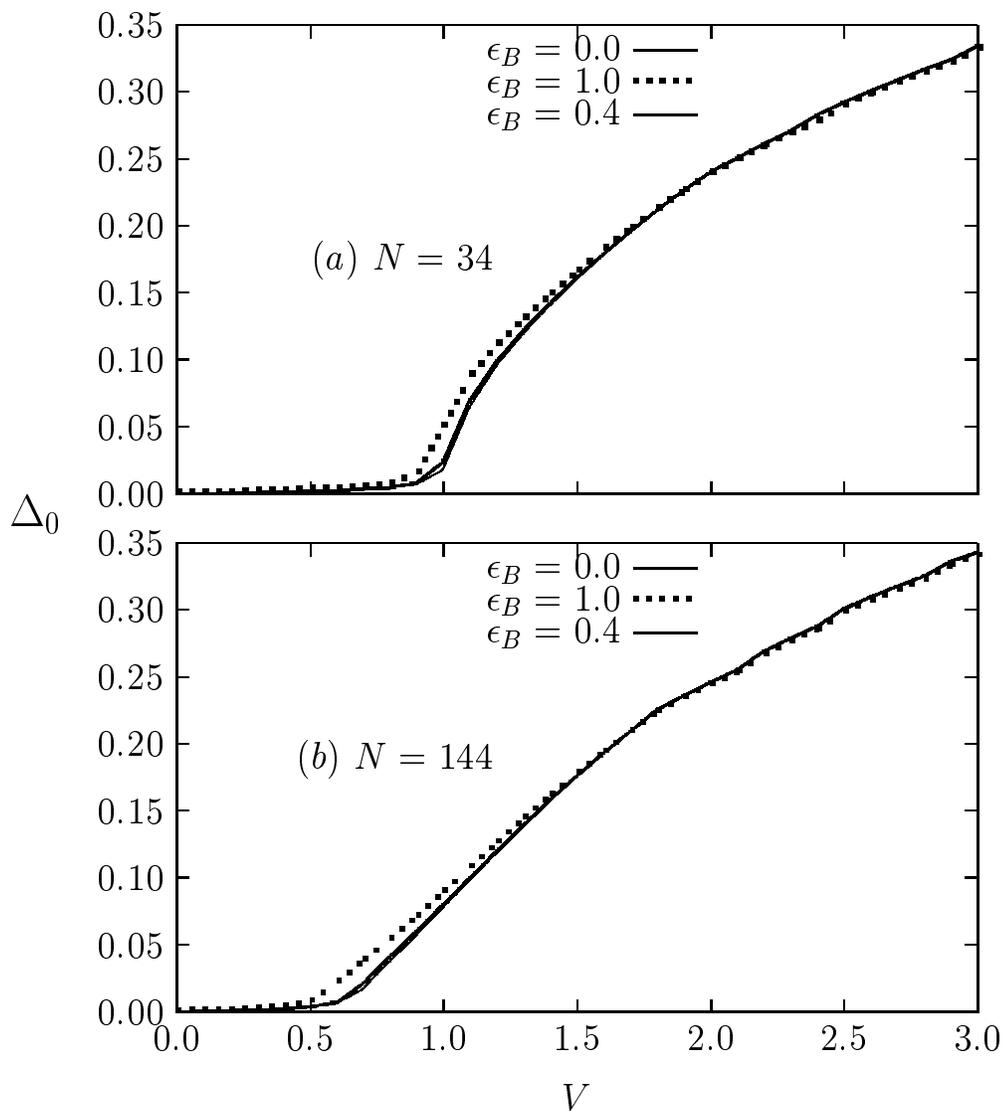
Fig.7. Plots of $\Delta_0$ vs. $V$ (positive $V$ sector) in the 
Fibonacci-modulated Penson-Kolb model for (a) $N=34$ and (b) $N=144$
for different strengths of disorder ($\epsilon_B$) (scale of energy is given by $t=1.0$).
\\
\\
\noindent
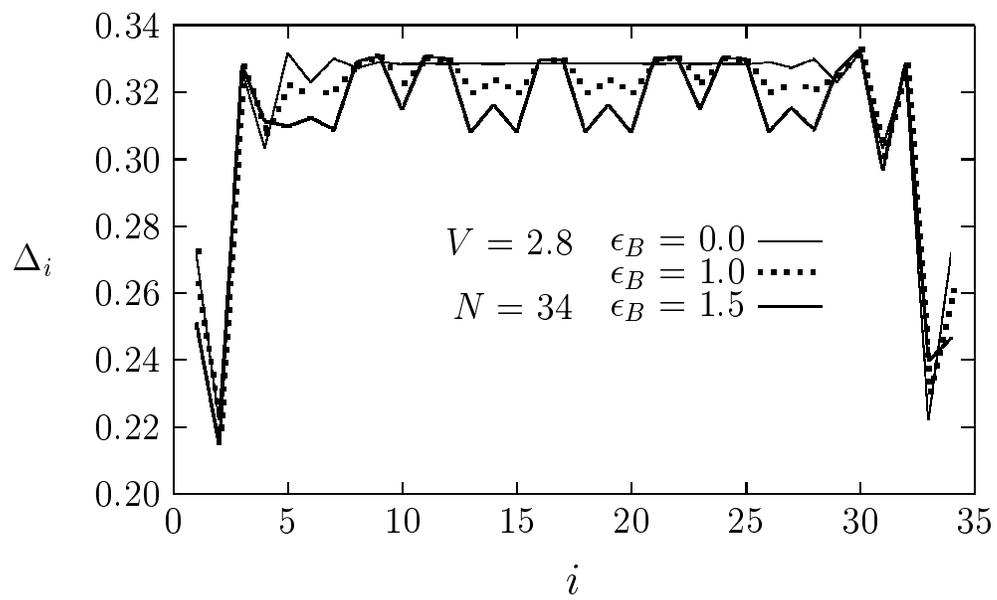
Fig.8. Plot of $\Delta_i$ vs. $i$ for the Penson-Kolb model (positive $V$
sector) with different strengths of Fibonacci ``disorder''. Fibonacci pattern
in $\Delta_i$ is very much suppressed compared to the case of the negative-$U$
Hubbard model (Fig.3) even for a reasonably large disorder ($\epsilon_B=1.5$) 
scale of energy is given by $t=1.0$).
\\
\\
\noindent
Fig.9. Plot of $\Delta_q$ vs. $q$ in the Fibonacci-modulated Penson-Kolb model 
for the negative $V$ sector for (a) $N=34$ and (b) $N=144$ for different strength 
of "disorder" ($\epsilon_B$) (scale of energy is given by $t=1.0$).
\\
\\
\noindent 
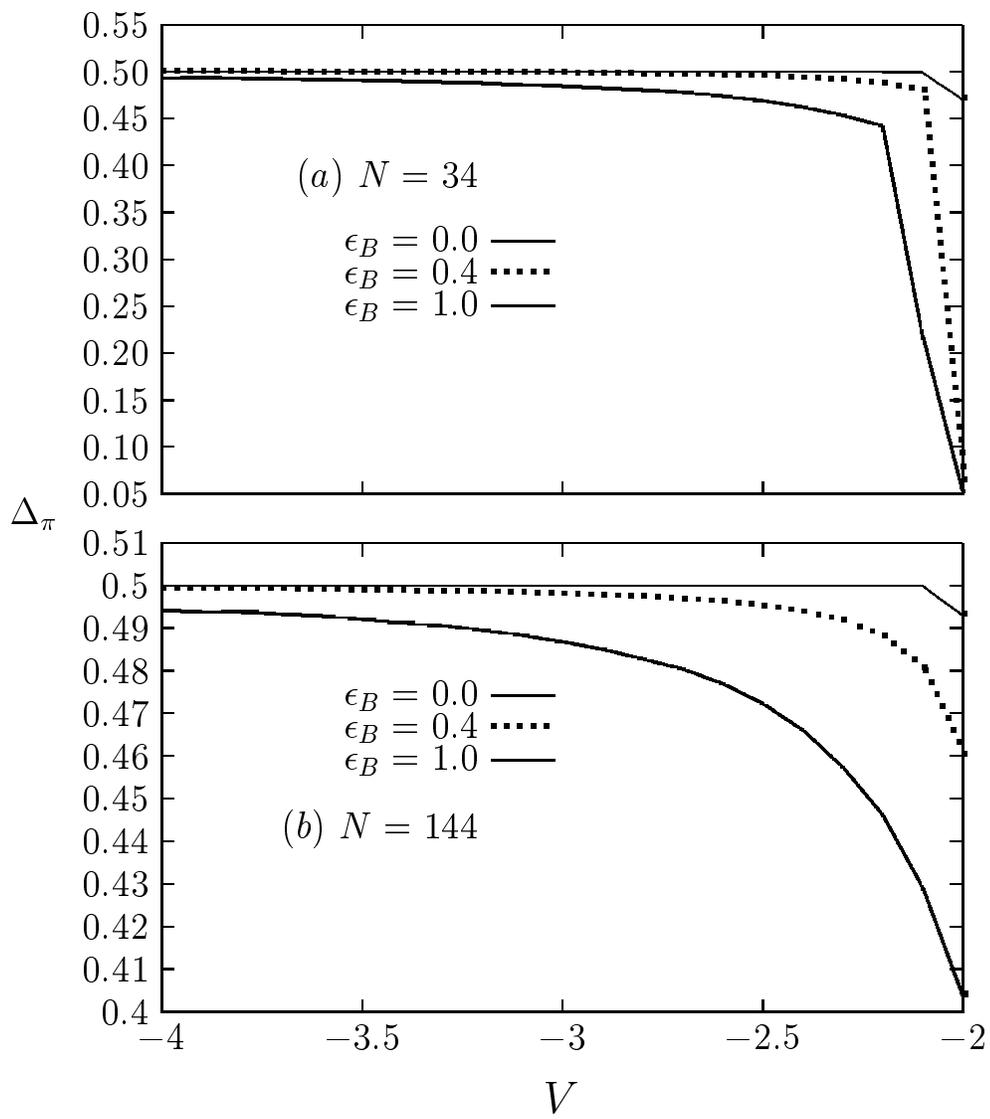
Fig.10. Plots of the superconducting gap $\Delta_\pi$ (for $\eta$ pairing 
center of mass momentum = $\pi$) vs. $V$ in the Fibonacci-modulated Penson-Kolb
model (in the negative $V$ sector) for (a) $N=34$ and (b) $N=144$. Effects of 
different strengths ($\epsilon_B$) of "disorder" are shown (scale of energy is given by $t=1.0$). 
\\
\\
\noindent
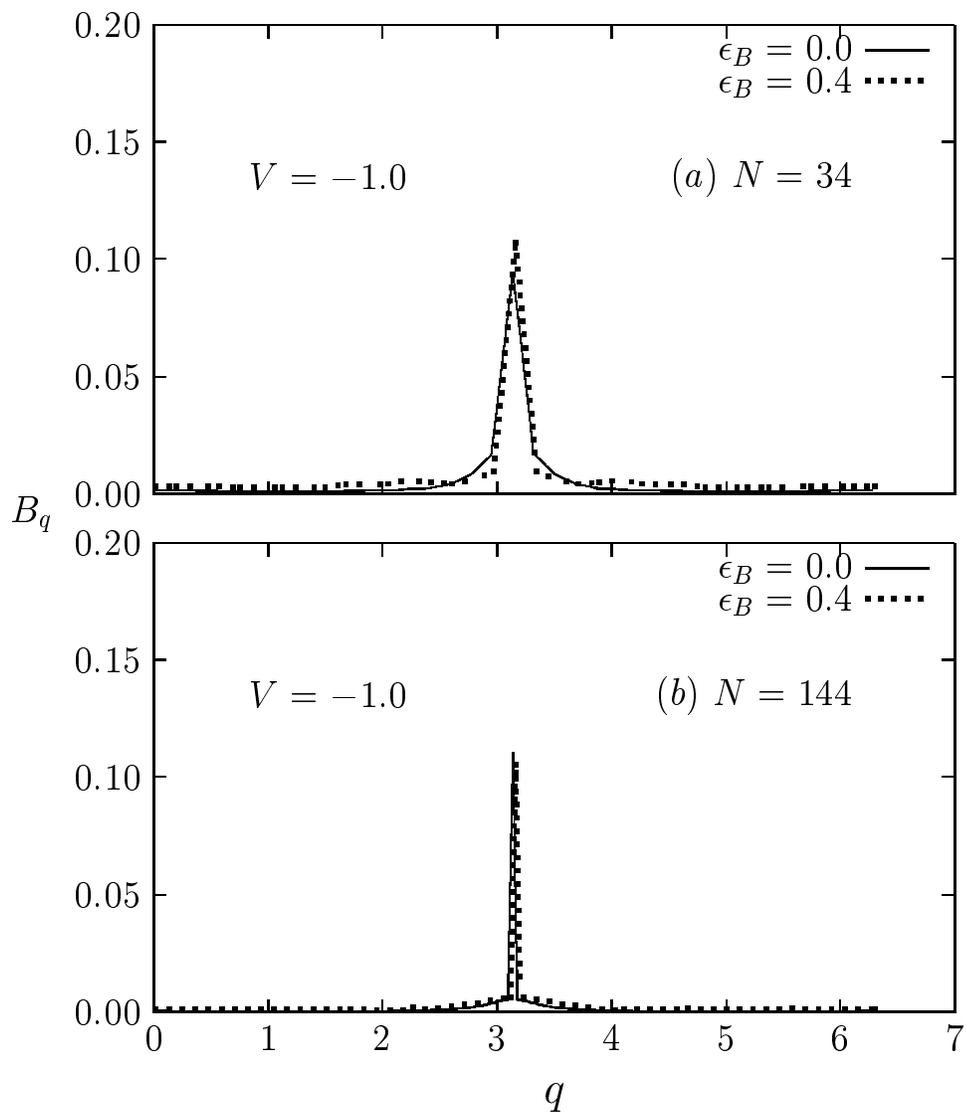
Fig.11. Plots of $B_q$, the Fourier transform of BOW order parameter, is 
shown as a function of $q$ for (a) $N=34$ and (b) $N=144$ for the Fibonacci-modulated 
Penson-Kolb model (scale of energy is given by $t=1.0$). 
\\ 
\\
\noindent  
Fig.12. Simultaneous plots of $B_{\pi}$ and $\Delta_{\pi}$ vs. $V$ 
in the negative $V$ sector of the Penson-Kolb model for (a) $\epsilon_B=0.0$ 
and (b) $\epsilon_B=0.4$ for $N=34$. Transition from BOW to $\eta$-paired phase 
is clearly visible and the transition point shifts towards larger values of $|V|$ for 
the increase of the  ``disorder'' strength (scale of energy is given by $t=1.0$). 
\\
\\
\noindent 
Fig.13. Simultaneous plots of $B_{\pi}$ and $\Delta_{\pi}$ vs. $V$ 
in the negative $V$ sector of the Penson-Kolb model for (a) $\epsilon_B=0.0$ 
and (b) $\epsilon_B=0.4$ for $N=144$. The transition point (corresponding to 
the transition from  BOW to $\eta$-paired phase) is virtually insensitive to 
the ``disorder'' strength (scale of energy is given by $t=1.0$).  
 
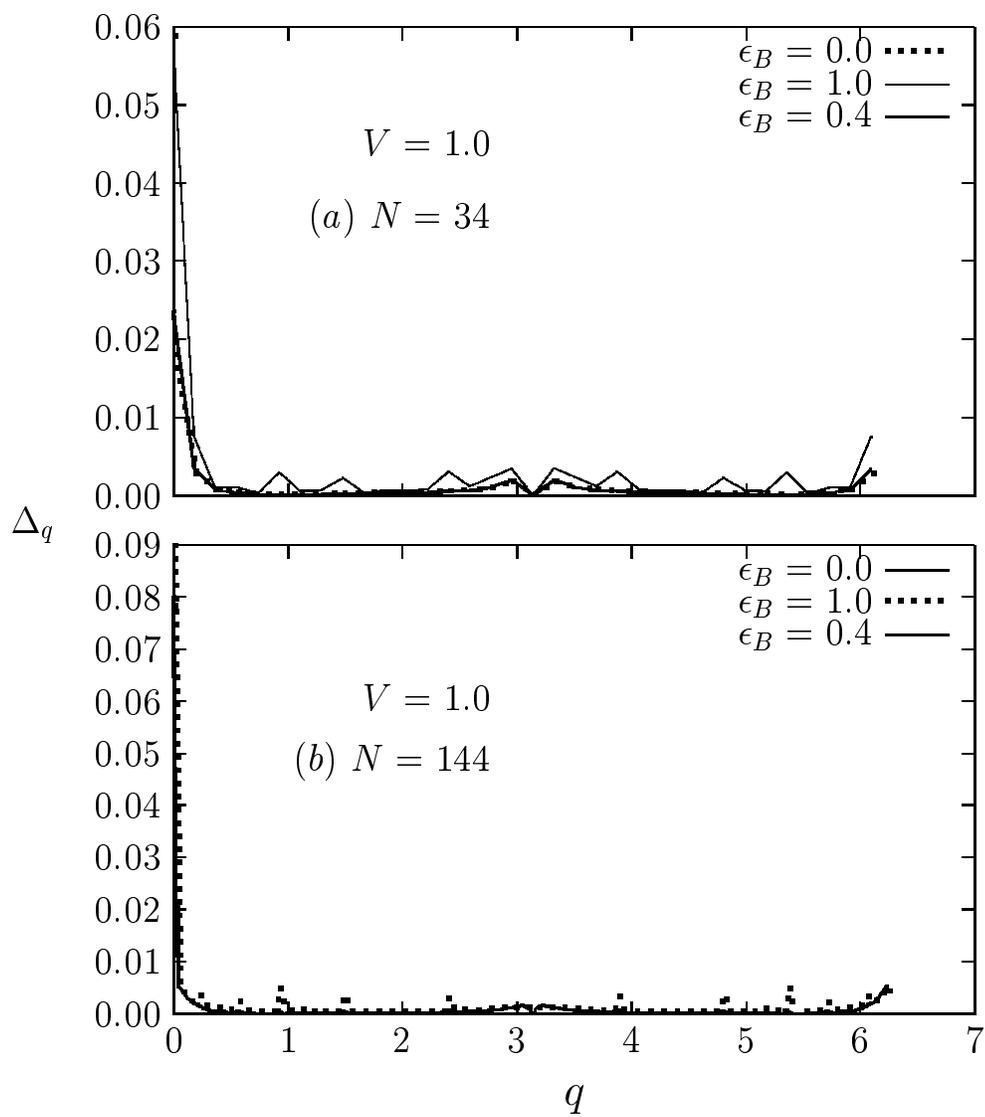
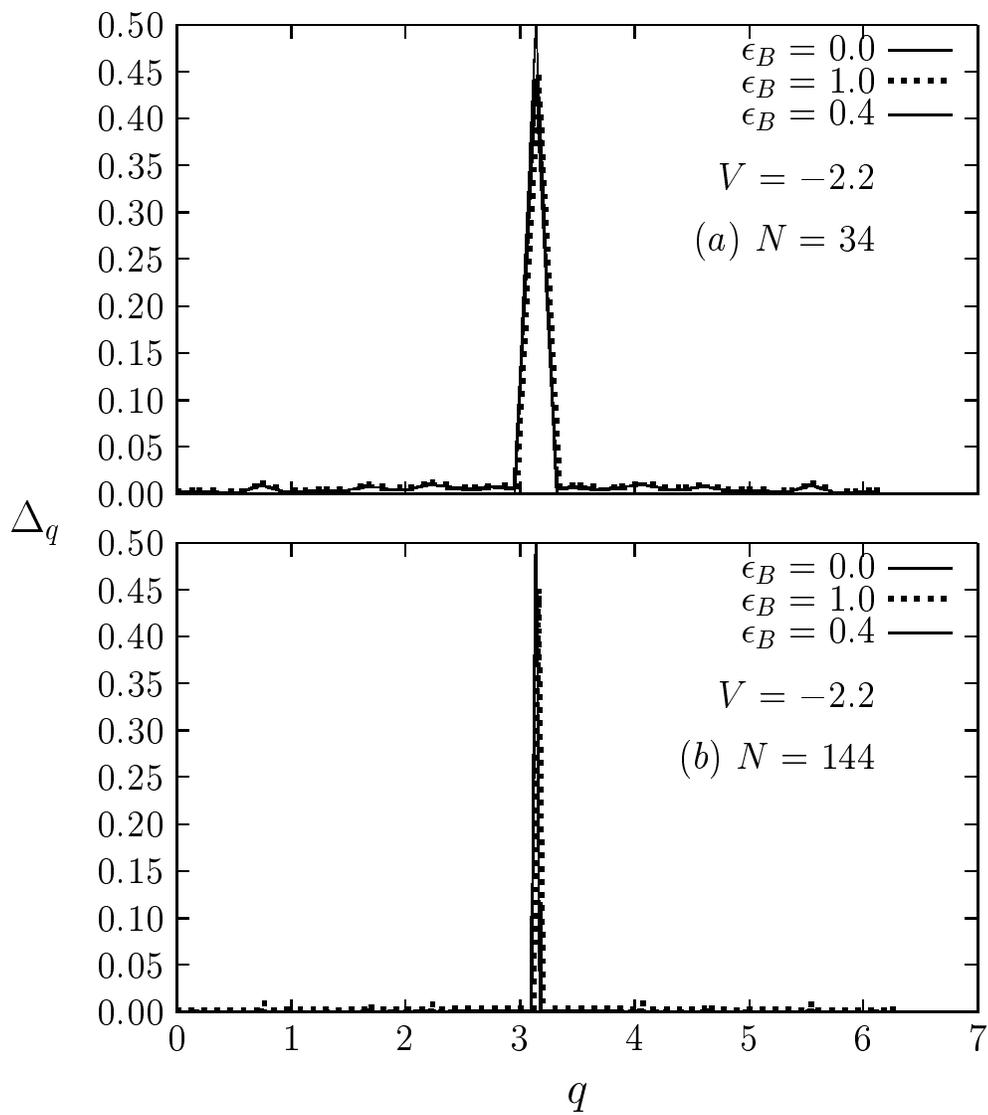
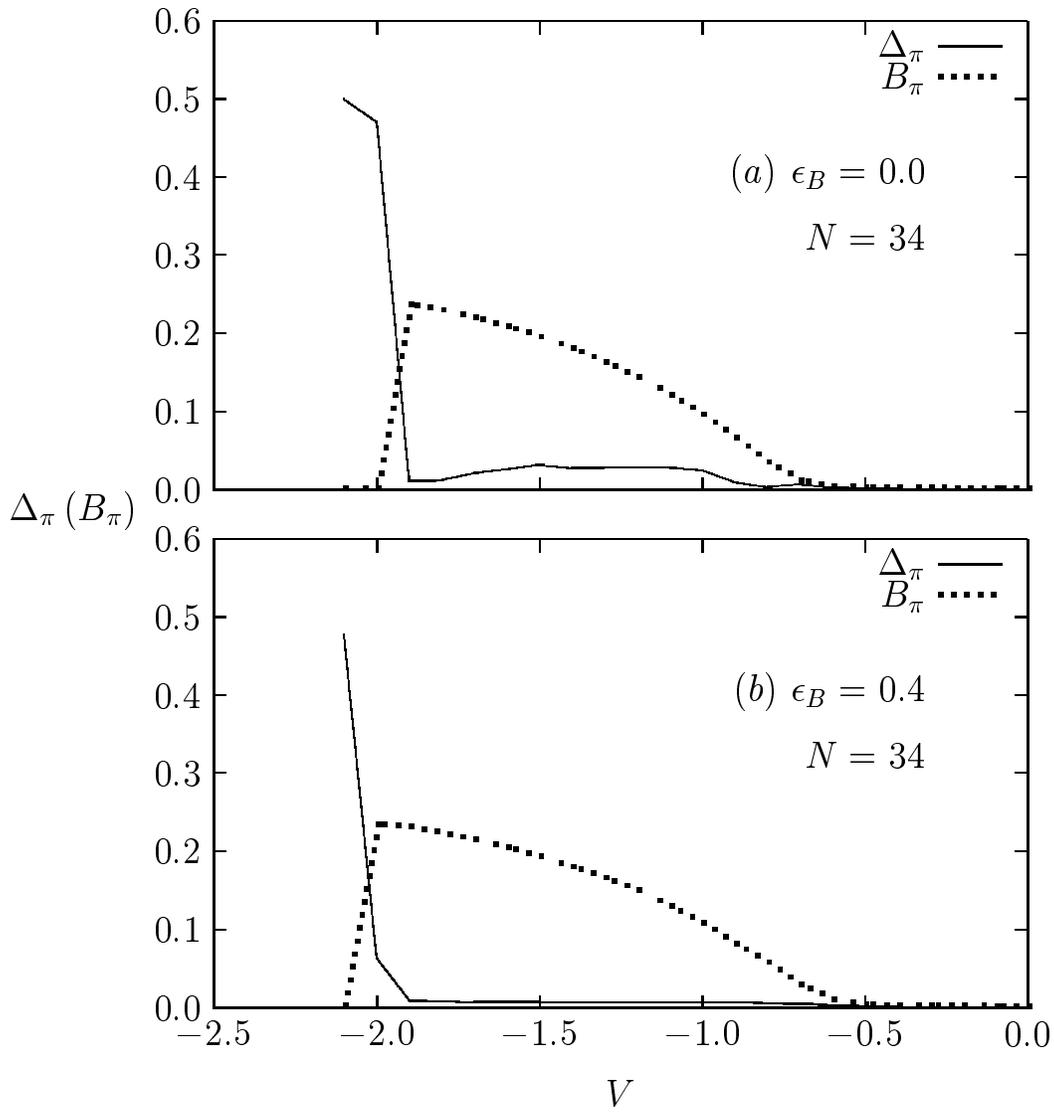
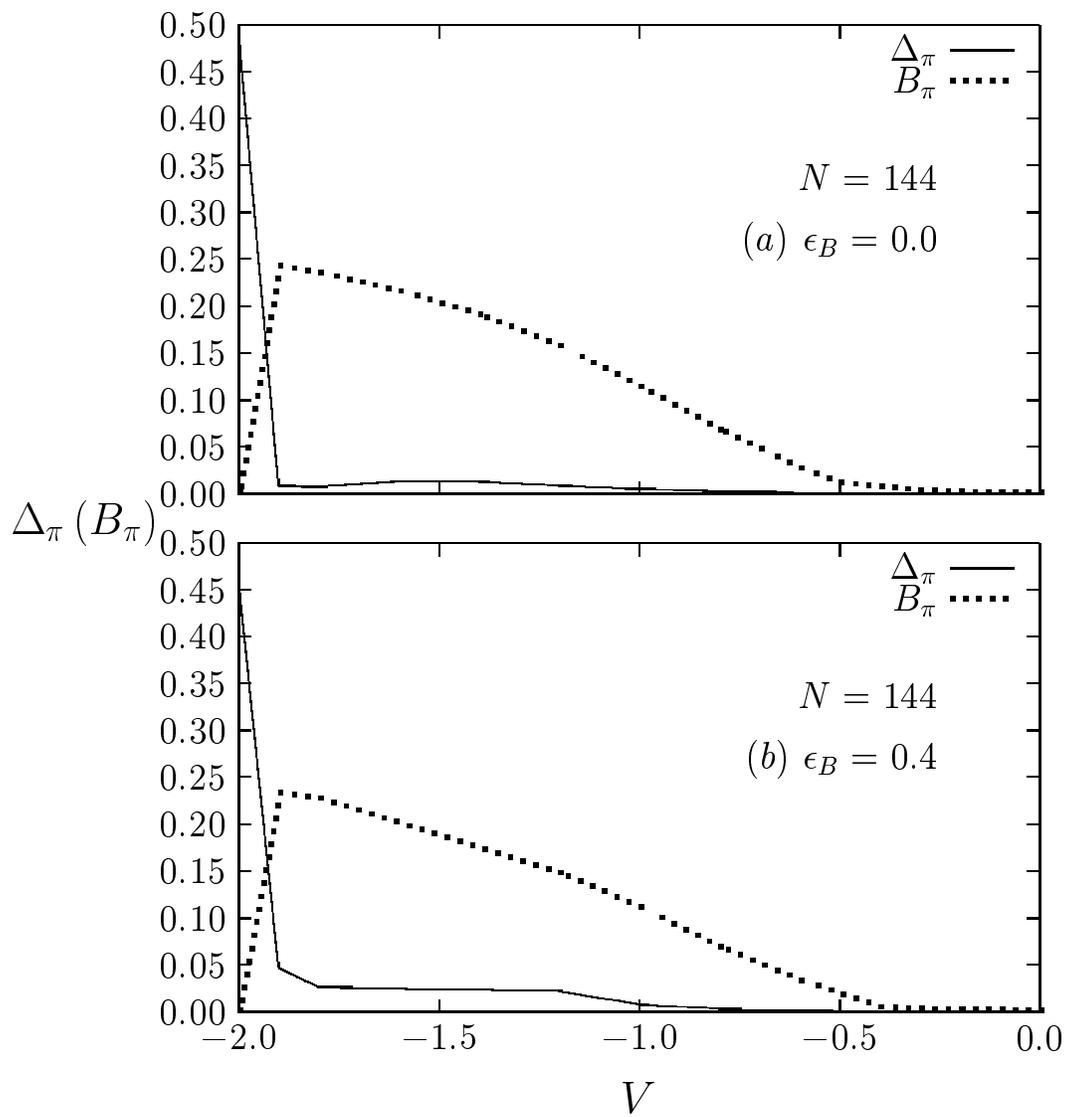
\end{document}